%% file: main.tex
\documentclass[review,authoryear]{elsarticle}

\usepackage{hyperref}
\usepackage{amsfonts}
\usepackage{amsmath}
\usepackage{graphicx}
\usepackage{wrapfig}
\usepackage{lscape}
\usepackage{rotating}
\usepackage{epstopdf}
\usepackage{booktabs}
\usepackage{pdflscape}
\usepackage{cite}
\usepackage{xcolor}
\usepackage{float}
\usepackage[margin=1in]{geometry}
\usepackage{setspace} 
\journal{Finance Research Letters}
\begin{document}

\begin{frontmatter}

\title{Russia’s Ruble during the onset of the Russian invasion of Ukraine in early 2022: The role of implied volatility and attention}

\author[sl]{\v{S}tefan Ly\'{o}csa}
\author[tp]{Tom\'{a}\v{s} Pl\'{i}hal\corref{cor1}}

\cortext[cor1]{Corresponding author \\ © 2022. This manuscript version is made available under the CC-BY-NC-ND 4.0 license\\ 
\url{https://creativecommons.org/licenses/by-nc-nd/4.0/}}

\address[sl]{Institute of Financial Complex Systems, Masaryk University, Lipova 41a, 602 00 Brno, Czech Republic; Faculty of Management and Business, University of Presov, Konstantinova 16, 080 01 Presov, Slovakia; Slovak Academy of Sciences, Sancova 56, 811 05 Bratislava, Slovakia; Email: stefan.lyocsa@gmail.com}

\address[tp]{Institute of Financial Complex Systems, Masaryk University, Lipova 41a, 602 00 Brno, Czech Republic; \\ Email: tomas.plihal@econ.muni.cz}

\begin{abstract}
The onset of the Russo-Ukrainian crisis has led to the rapid depreciation of the Russian ruble. In this study, we model intraday price fluctuations of the USD/RUB and the EUR/RUB exchange rates from the $1^{st}$ of December 2021 to the $7^{th}$ of March 2022. Our approach is novel in that instead of using daily (low-frequency) measures of attention and investor's expectations, we use intraday (high-frequency) data: google searches and implied volatility to proxy investor's attention and expectations. We show that both approaches are useful in predicting intraday price fluctuations of the two exchange rates, although implied volatility encompasses intraday attention.
\end{abstract}

\end{frontmatter}

\section{Introduction}
The tension between Russia and Ukraine began to increase in February 2014, following the Ukrainian Revolution of Dignity, also known as the Maidan Revolution. As Ukraine was facing internal conflicts, Russia invaded and subsequently annexed the Crimean Peninsula in February and March 2014, which is territory that is still (as of March 2022) internationally recognized as part of Ukraine. Following these events, in March 2014, a war also began in Donbas (the Donetsk and Luhansk regions of Ukraine), where pro-Russian anti-government separatists who were supported by Russia led an armed conflict against the Ukrainian government forces \citep{mankoff2014russia, rywkin2014ukraine, shelest2015after, samokhvalov2015ukraine}.

After nearly eight years, the conflict has not been resolved. Instead, it has escalated. During November and December 2021, Russian troops started to increase their presence near Ukrainian borders (also in Belarus and Crimea), which they masked as regular military exercises. On February 21, 2022, Russian President Vladimir Putin announced that Russia recognized the independence of two pro-Russian regions, namely, Donetsk and Luhansk, in eastern Ukraine. It triggered the first round of economic sanctions from NATO countries in the following days. The whole situation escalated on February 24, 2022, when Russia began a complete invasion of Ukraine, which led to a full-fledged war that surprised most of the global population. As a result, the economic sanctions from other countries against Russia become even more severe, including restrictions to Russian imports and exports, the removal of selected Russian banks from the SWIFT interbank system, and the prohibition of the Central Bank of Russia from access to foreign exchange reserves, to name a few\footnote{A comprehensive list of all sanctions can be found at \newline https://graphics.reuters.com/UKRAINE-CRISIS/SANCTIONS/byvrjenzmve/}.

The following Figure 1 shows how the combination of war and economic sanctions has impacted the Russian ruble. The ruble significantly weakened and lost almost 50\% of its value against the USD over the course of a few days at the end of February and the beginning of March in 2022. Volatility is a key parameter for valuing assets with uncertain future payoffs and in such an uncertain crisis period, an accurate volatility model is in demand by institutional investors and policymakers alike, as it can be used to understand the value of assets that are tied to the value of the Russian ruble. Based on the existing literature, we hypothesize that we can create a suitable prediction model based on i) limited attention theory and ii) the forward-looking nature of option contracts. More specifically, in this study, we are interested in predicting intraday price fluctuations of the ruble against the USD and the EUR during the onset of the Russo-Ukrainian crisis using the population's attention and investors' expectations.

Our research is related to several strands of the literature. First, we contribute to the emerging literature on the effects of the onset of the Russo-Ukrainian war on the economy and markets by analyzing the period from the $1^{st}$ of December 2021 to the end of the $7^{th}$ of March 2022. \citet{mamonov2021sorry} examined how significant the macroeconomic effects of financial sanctions are using the Bayesian (S)VAR model. They focused on the Western financial sanctions that were imposed on the Russian economy in 2014 and 2017. The effects of these sanctions were negative and non-negligible. The most significant impact was found on the real interest rate and external corporate debt, while a moderate impact was found on the ruble real exchange rate, output, consumption, investment, and trade balance. Recently, \citet{liadze2022economic} estimated and evaluated possible economic costs of the Russia-Ukraine conflict, while \citet{ozili2022global} analyzed the global economic consequences of the Russian invasion of Ukraine, the effect of sanctions imposed on Russia, and spillover effects relevant to the global economy. \citet{halouskova2022} recently showed that the amount of attention paid to the Russo-Ukrainian crisis period could be used to predict the next day's price fluctuations in stock markets around the world; the authors also found that geographical and economic distance is negatively related to the relationship between attention and price fluctuations. Finally, a related study to our research is \citet{polyzos2022escalating}, who used Twitter to extract number of tweets (an attention measures) and a sentiment index to study their impact on intraday returns of several stock market indices, commodities, the U.S. Treasury bill index, Bitcoin and three currencies, including Ruble. Using a vector autoregression framework, \citet{polyzos2022escalating} provides evidence that shocks in the number of tweets were related to the depreciation of the Russian Ruble.

Second, we contribute to the literature on limited attention theory \citep{barber2008all} that uses Google Trends data to predict future price variation \citep{goddard2015investor,dimpfl2016can,audrino2020impact,lyocsa2020impact}. Our idea to explain price fluctuations via attention measures is based on the limited attention theory of \citet{barber2008all}, who argue that all investors are subject to a limited amount of information they can process. Thus, information that receives the most attention is likely to influence investor's behavior the most. We hypothesize that limited attention theory is particularly relevant for the Russo-Ukrainian war that has attracted considerable attention from the worldwide media. Attention itself does not cause trading, but it is a proxy of an investor's interest in a particular news topic, and information arrival is known to be related to market activity and volatility in particular \citep[e.g.][]{melvin2000public,plihal2021scheduled}. Attention can be estimated from different sources, like news articles, Wikipedia page reads, Twitter, or Google search volume indices. The latter reference is considered to be a good proxy for information demand \citet{bleher2021knitting}, as Google searches are highly popular in many parts of the world. In this study, we also rely on Google searches that are well-established in the finance literature \citep[e.g.][]{da2011search} and have been successfully used in previous studies to model future price variations \citep[e.g.][]{dimpfl2016can,audrino2020impact,lyocsa2020impact}. We extend the current literature in that we address the question of whether limited attention theory manifests for intraday trading as well. Using the intraday measures of attention, which is unique in the literature, we show that high-frequency attention is still valuable for predicting fluctuations in intraday prices.

\begin{landscape}
\begin{figure}
  \centering
  \caption{Price of the USD/RUB exchange rate during the onset of the Russo - Ukrainian war}
  \label{figure1}
  \includegraphics[width=8in]{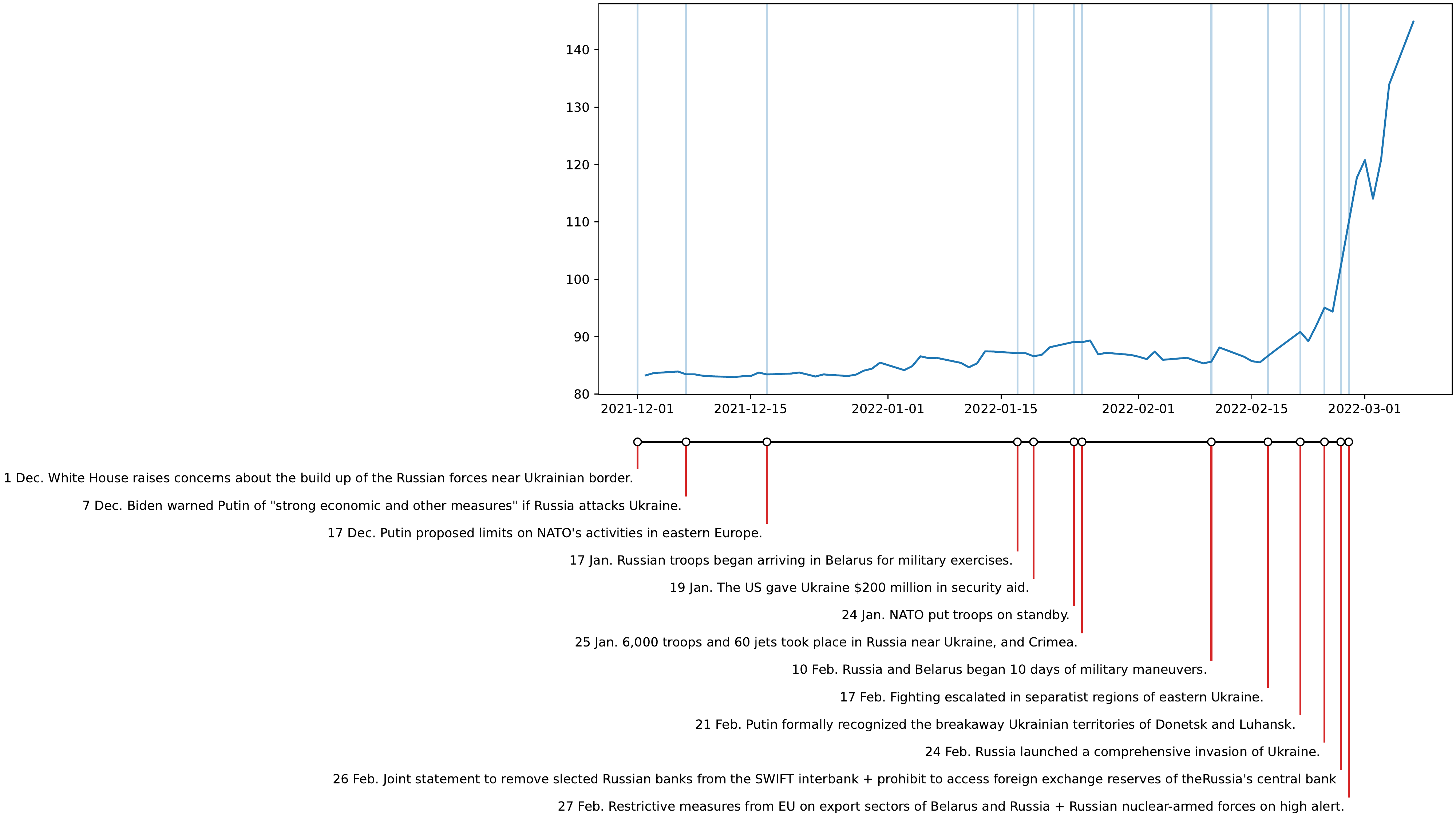}
\end{figure}
\end{landscape}

Our third contribution is related to the implied volatility literature. Market participants enter option contracts based on their expectations about the future outcome of the underlying assets over the course of the contract's maturity. Therefore, implied volatility, derived from option contracts, should contain all the available information relevant to future price fluctuations in an efficient market. In a recent study, \citet{fassas2021implied} studies implied volatility indices across multiple asset classes, including currency implied volatility indices. Within currency indices, \citet{fassas2021implied} report a negative relationship between implied volatility and returns and a positive relationship between realized and implied volatility. If both implied and realized volatilities are persistent, results of \citet{fassas2021implied} suggest that implied volatility might be useful in predicting future levels of realized volatility. A strand of the literature seems to exploit the link between implied volatility's information content and future realized volatility \citep{canina1993informational, poon2003forecasting,gonzalez2015model}. For example, the recent study of \citet{plihal2021modeling} showed that implied volatilities are more useful when modeling and predicting future realized price variations of the EUR/USD currency pair than historical price variations. Our approach is novel in that we address the question of the utility of intraday implied volatility, which to the best of our knowledge, is also a unique approach. However, there are several reasons why implied volatility might not be as informational in our setting. First, it is known that implied volatilities tend to be biased upwards, which was likely amplified during the onset of the Russo-Ukrainian war as relevant information was unfolding gradually from one hour to the next. Therefore, options were likely priced with an even higher risk premium. Second, as opposed to the EUR/USD market, the USD/RUB and the USD/EUR are smaller markets that might not be as efficient. Third, our analysis is based on intraday implied volatilities that have not been used in previous studies and are likely noisier proxies for future price expectations. We also show that intraday investors' expectations drive intraday realized price fluctuations. Finally, we acknowledge that attention and implied volatility might be related, which leads us to compare their role in predicting price fluctuations. Consequently, we provide strong evidence that the implied volatility likely encompasses investors' attention. 

The paper is organized as follows. The next section describes the data sources and the construction of key variables of interest. The following section specifies our empirical strategy. The presentation of the results and a related discussion follows, while the final section concludes the paper and discusses a future research agenda. 

\section{Data and measures}\label{Data}

\subsection{Exchange rate data and price-variation}

Russian troops mobilized near the Ukrainian border during October and November of 2021, and in early December, this threat was publicly recognized by the White House\footnote{see an article from the $1^{st}$ of December 2021,  https://www.theguardian.com/world/2021/dec/01/us-warns-russia-plans-large-scale-attack-o n-ukraine.}, which attracted considerable attention by the media. Our sample of data, therefore, starts on the $1^{st}$ of December 2021 and ends on the $7^{th}$ of March 2022, which was the most recent date available prior to the start of this research. We use $5-$minute trading prices on two exchange rates, namely, the USD/RUB and the EUR/RUB, from Bloomberg, which were retrieved in UTC (Coordinated Universal Time) for whole 24-hour periods spanning from Monday to Friday.

In order to estimate price reactions to attention and investor's expectations (implied volatilities), we divide the $24$ hour foreign exchange trading day into six consecutive $4$ hour trading windows. We thus estimate and consequently model intraday price variation, realized variance, within a $4-$hour trading window, as given by the corresponding realized variance. Let $t = 1, 2, ..., T = 406$ denote the usual time index, now corresponding to consecutive $4-$hour trading windows. Within each trading window, we have $j = 1, 2, ..., J = 48$ price observations, i.e. $4$ hours and $12$ prices sampled after $5$ minutes.\footnote{The 5-minute sampling frequency is usual in the literature, e.g. \citet{liu2015does,gkillas2021uncertainty}. \citet{liu2015does} has provided empirical evidence that it is difficult to out-perform 5-minute sampling frequency, although there were instances where 1-minute sampling frequency was doing well. For a recent application of the 1-minute sampling frequency see \citet{gkillas2021uncertainty} .} Now let $P_{t,j}$ denote the exchange rate, the annualized realized variance for a given trading window $t$ is given by:
\begin{equation}
    V_{t} = 252 \times 6 \times \sum_{j = 2}^J \left[100 \times \left(ln(P_{t,j}) - ln(P_{t,j-1})\right)\right]^2
\end{equation}

\subsection{Implied volatility data}

We capture investors' anticipations through implied volatility measures. Similarly to \citet{plihal2021modeling}, we use implied volatilities that are directly traded mostly by institutional investors (banks and insurance companies). Investors trade currency options (over-the-counter market), through the Bloomberg terminal, by quoting implied volatilities directly. The advantage to this approach is that one does not need to calculate implied volatilities from option prices while assuming a specific option pricing model; instead, we use quoted implied volatilities directly.
In our empirical application, we retrieve already annualized implied volatilities\footnote{Implied volatilities are already quoted in an annualized form in order to allow comparison across different option maturities.} as given by the USD/RUB and EUR/RUB options traded at-the-money (ATM), which have the same delta for a monthly maturity. We use $5-$minute data from the UTC timezone. The implied volatility averaged across the $4-$hour window is given by the following:
\begin{equation}
    IV_{t} = J^{-1} \sum_{j = 1}^J I_{t,j}^2
\end{equation}
, where $I_{t,j}$ is the quoted volatility in trading window $t$ and for $j^{th}$ $5-$minute sample. In order to have implied and realized volatilities on the same scale, we first take the square of the quoted volatility in the equation above. Note, that we also consider daily, weekly, two-week, three-week and three-month maturity options, but as the values are correlated significantly, we use the monthly maturity as is usual in the implied volatility literature.

\subsection{Attention measures}

We use Google Trends as a measure of attention, which is consistent with the increasing number of studies that suggest that attention might drive stock market activity\footnote{For recent applications to financial markets, see \citet{kim2019google,bleher2019today,lyocsa2020fear,aslanidis2022link}.} (i.e., volume, volatility, extreme price movements). Google publishes search volume indices that represent the relative popularity of a search term(s) within a given time period and across multiple search terms. According to their construction, these search volume indices are rounded to integers and bounded to $1$ and $100$, where $100$ corresponds to the highest popularity within the given sample period and across multiple search-terms\footnote{Note that we restrict the lowest possible value of the search volume index to be 1, not 0, which is introduced for mathematical convenience.}. To capture the population's attention, we retrieve multiple terms related to the events surrounding the Russo-Ukrainian war. Each term is retrieved separately with no geographical restriction (i.e., the whole world) and for English language searches, using the package of \citet{massicotte2021gtrendsr}. The use of Google Trends data is not straightforward; therefore, careful handling of the data is necessary.

First, as argued by \citet{chan2020temporal,bleher2021knitting}, in one batch, Google Trends allows one to retrieve a maximum of $270$ consecutive observations that are scaled from $1$ to $100$. Concatenating these series into one group (which is unfortunately often used) would lead to jumps in the break-points because of the different scales of each batch, see \citet{bleher2021knitting}. A better approach might be to employ logarithmic differences before concatenation; however, \citet{bleher2021knitting} shows that this approach can also lead to changes in the underlying distributional properties. \citet{bleher2021knitting} recommends using a regression-based procedure, and another viable alternative might be the procedure used by \citet{chronopoulos2018information}. Our approach is more similar to that of \citet{bleher2021knitting} and even more so to that of \citet{kristoufek2015main} as we do not use a regression. The similarity lies in the fact that for each term, we download search volume indices for each trading window\footnote{We retrieve google search volume indices for each hour, and we use the average of the $4-$hour trading window, i.e. over $4$ observations, to arrive at the search volume index for the given trading window $t$.} $t$ in a batch covering $4-$days, where consecutive $4-$day batches overlap across $1$-day, i.e. one day overlaps with the previous and one day overlaps with the next batch (except the first and last batch). The overlapping batches are used to re-scale and chain the data\footnote{The following example illustrates our approach. Let $S_t(2)$ denote the search volume indices downloaded in batch 2, with specific realizations being $20, 100, 18, 14, 28, 26, 15, 14, 12, 10$ for periods, $t = 8, 9, 10 ,11, ..., 17$, and let $S_t(1)$ represent the search volume indices in batch 1, with realizations of $50, 60, 70, 80, 100, 40, 30, 5, 25, 5$, for periods $t = 1, 2, 3, ..., 8, 9, 10$. The two batches overlap in periods $t = 8, 9, 10$. Note that taking the ratio within the overlapping volumes leads to following re-scaling constants: $S_{t=8}(b=2) / S_{t=8}(b=1) = 4, S_{t=9}(b=2) / S_{t=9}(b=1) = 4, S_{t=10}(b=2) / S_{t=10}(b=1) = 3.6$. Ideally, these should be the same, in which case the re-scaling is straightforward and only a small overlapping period would be necessary. However, this is often not the case due to the rounding of integers. This rounding effect might lead to uncertainty related to the re-scaling constant, and importantly, it is amplified for smaller search volume indices. We recommend using batches with considerable overlap to mitigate the rounding effect and, consequently, to re-scale the series by taking the maximum of the estimated re-scaling constant. Thus in the previous example, we would use a constant of $4$, and the resulting series would be $200, 240, 280, 320, 400, 160, 120, 20, 100, 18, 14, 28, 26, 15, 14, 12, 10$.}. The resulting search volume index for a given term is denoted as $A_t(m)$, where $m \in M$ is a specific term. 

Second, Google Trends are also retrieved for weekends (or non-trading days in general). If we are interested in predicting price variation for the first trading window on Monday, i.e., from 00:00 to 04:00 UTC Monday, then the lagged attention corresponds to Sunday from 20:00 to 00:00 and not Friday.

Third, we need to specify search terms. We use three categories: i) general financial market (e.g., SP 500, VIX, FX market,...), ii) ruble (e.g., ruble, USD rub, Russian interest rate,...), and iii) Russian economy-related (e.g., economic sanctions, asset freeze, Nord Stream 2, export controls, British Petroleum Russia, Ikea Russia,...) search terms. Within each category, we have sets of search terms ($M_G, M_R, M_E$, see Appendix A). Let $|.|$ denote the cardinality of a set; the final series of attention toward general financial markets is given by the following:
\begin{equation}
  G_t = |M_G|^{-1} \sum_{g \in M_G} A_t(g)
\end{equation}
The same approach is applied for the ruble $R_t$ and economic sanction $E_t$ attention series.

\section{Methodology}\label{method}

In order to model the intraday price variation of the Russian ruble to the U.S. dollar and the euro, we need to account for the serial dependence of the volatility series and the possible intraday seasonality introduced by $24-$hour weekday trading and different liquidity and trading activities over the trading day. As noted in the previous sections, a trading day is split into six trading windows, i.e., $j = 1, 2, ..., J = 6$. We introduce a dummy variable $I_t(j)$ that takes a value of 1 if the next period of variation belongs to trading window $j$ and $0$ otherwise. Note that $j = 1$ corresponds to the trading period from 00:00 to 04:00 and excluded trading period, $j = 6$, is the one ranging from 20:00 to 00:00. 

As the conflict in Ukraine has escalated, the amount of exchange-rate-relevant news has increased considerably, thereby making it impossible to account for specific effects. For example, to stabilize its currency, the Central Bank of Russia announced an increase of the baseline rate to $20\%$ on the $28^{th}$ of February 2022. However, at the same time, Directive 4 of the U.S. Office of Foreign Asset Control under Executive Order 14024 was introduced, which prohibits any transactions with key financial institutions in Russia, including the Central Bank of Russia. To account for this general increase of trading-related news, our specifications also include the days that occurred during the onset of the war, namely, the trading days ranging from the $21^{th}$ to the $28^{th}$ of February 2022. The dummy variable $I_t(q)$ takes a value of 1 if the next given intraday period price variation falls to the day $q$ and $0$ otherwise.

An established stylized fact about volatility is that it is a highly persistent time series and does not contain a unit root (e.g. \citealt{andersen2001distribution}). Given a limited sample period and size surrounding the event of interest, we opted for an autoregressive volatility model, resembling the heterogeneous autoregressive (HAR) model of \citet{corsi2009simple}\footnote{An alternative is to use a generalized autoregressive conditional heteroscedasticity (GARCH) model of \citet{bollerslev1986generalized}. Given the limited sample size the maximum likelihood estimation of an overparametrized GARCH class model is not recommended.}. The standard HAR model uses daily, weekly, and monthly average volatilities to model future levels of volatility. However, we opt for a different specification that suits our research design more. For example, our sample period covers only around four months of data; thus, monthly and weekly volatility patterns are unlikely to emerge and are consequently not employed. HAR models are mostly employed for studying daily level of volatility, while we model intraday price fluctuations. In such scenario, existing literature suggests that intraday seasonality patterns influence the volatility persistence (e.g., \citealt{ederington2001intraday,chang2003information,seemann2011intraday}) and we control for their existence. Also volatility is susceptible to extreme observations, that we observed during the actual start of the military conflict in February. Such observations are likely to bias the parameter estimates. Therefore, we control for such abrupt changes in the market conditions by introducing specific day effects. The onset of the crisis is likely to cause a non-linear response in the market behavior, which led us to estimate the model in the log-log form. Finally, because of the high persistence level and limited sample size, we run a battery of unit-root panel tests (see description under Table 1); the results suggest that a unit root is not present in our series. However, during a time period of such extreme uncertainty, implied volatility might appear (in a finite sample) to have a unit-root-like behavior. Thus, to err on the safe side, we also consider a specification using the first differences of implied volatilities. We use seven model specifications, with the most general (Model 7; see Tables 3 and 4) being as follows:
            \begin{equation}
            \begin{split}
            ln V_{t} = & \beta_0 + \beta_1 ln V_{t-1} + ln V_{t-1} \sum_{j=1}^{5} \gamma_{j} I_t(j) + ln V_{t-1} \sum_{q=1}^{6} \lambda_{q} I_t(q) + \\
             & \eta_1 ln G_{t-1} + \eta_2 ln R_{t-1} + \eta_3 ln E_{t-1} + \pi_1 \Delta ln IV_{t-1} + \pi_2 ln IV_{t-1} + \epsilon_t
            \end{split}
            \end{equation}
As we estimate our model in the log-log form, the coefficients correspond to the elasticities of the given variables in relation to the underlying realized variance. In regard to interaction terms, $\gamma_j ln V_{t-1} I_t(j)$ corresponds to the elasticities of the given variables in relation to the underlying realized variance. In regard to interaction terms, $\gamma_j ln V_{t-1} I_t(j)$ corresponds to the intraday seasonality, while $ln V_{t-1} \lambda_q I_t(q)$ corresponds to specific day effects; furthermore, $\Delta ln IV_t$ corresponds to the first difference of $IV_t$. Note that the specification defined above is predictive in nature, as future price variation is explained by lagged values only. The model is estimated via OLS, and the significance of coefficients is derived via the stationary bootstrap method with a block length randomly drawn from the geometric distribution of the expected value, as suggested by the procedure of \citet{politis2004automatic, patton2009correction, Racine2021}. After each model, we check for the presence of the serial dependence of residuals using the test found in \citet{escanciano2009automatic}.

\section{Results}\label{results}

\subsection{Exploratory observations}

In Figure 1, we observe how the Russian ruble weakened after fighting escalated in the separatist regions of eastern Ukraine, which foreshadowed the upcoming invasion reaching deeper into the Ukrainian territory. This weakening was accompanied by increased price fluctuations of and attention to the ruble, as indicated in Figure 2. Despite the somewhat erratic behavior of the intraday price fluctuations and attention, Figure 2 suggests a remarkable synchronization of the related attention and price fluctuations after the onset of the Russo-Ukrainian war.

\begin{figure}
  \centering
  \caption{Price variations of and attention toward the Russian ruble}
  \label{figure2}
  \includegraphics[width=6.5in]{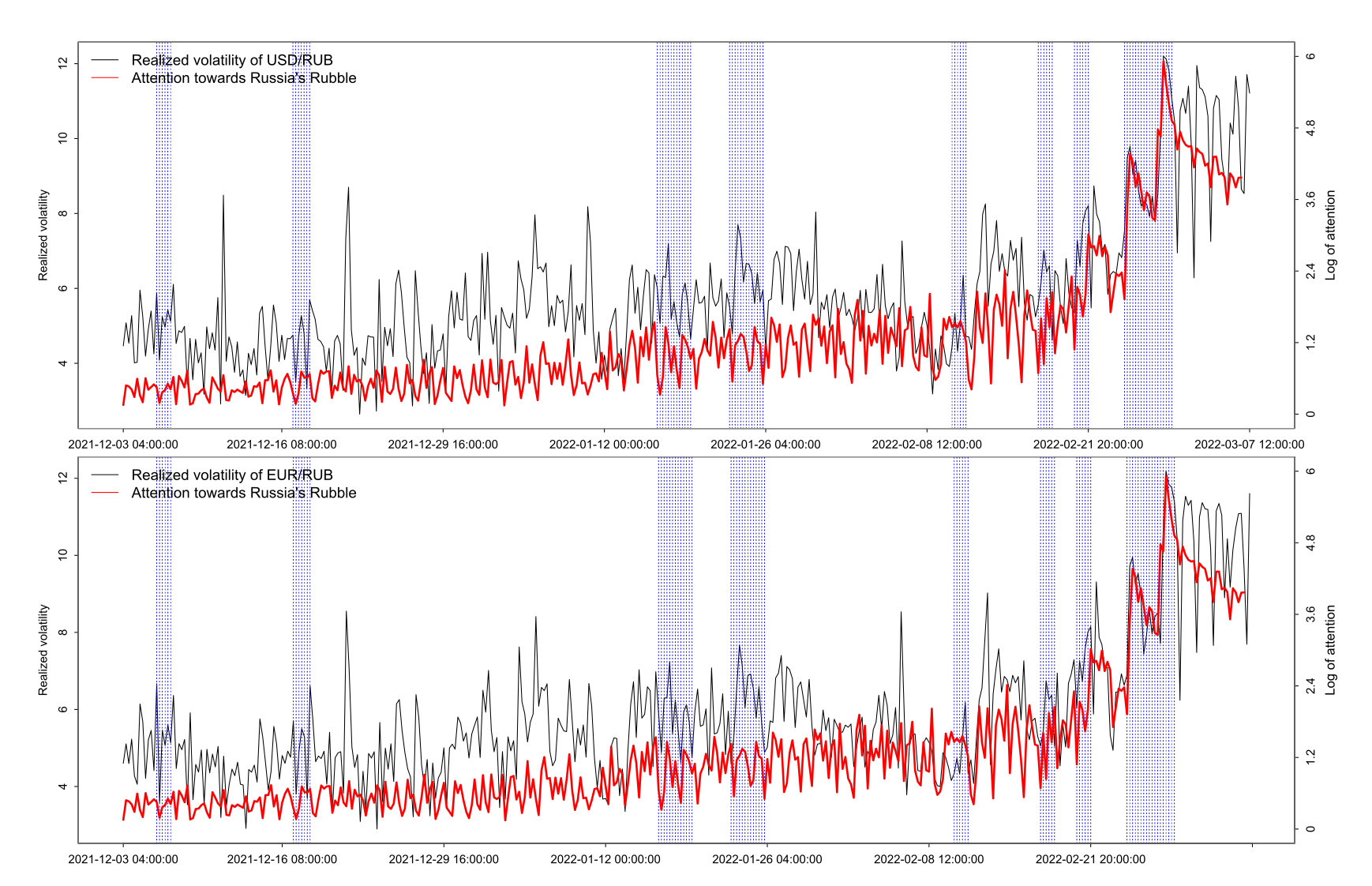}
\end{figure}

Figure 2 also shows that attention seems to be less erratic than the price fluctuations, as confirmed by the results shown in Table 1; furthermore, attention is shown to have declined after the onset of the war, while price fluctuations remain quite extreme. To illustrate the extreme levels of volatility of the USD/RUB exchange rate, consider that the annualized volatility across commodities, equities, foreign exchange and fixed income instruments (using daily data) is around $25.4$, $20.6$, $10.6$ and $3.1$ \citet{bollerslev2018risk}, respectively, while it ranges from $60$ to $100$ for Bitcoin \citet{nicholas2021bitcoin}. Over the observed period, the annualized volatility of both the USD/RUB and EUR/RUB exchange rates is around $19$ (calculated from $5.88$ for the USD/RUB exchange rate, as shown in Table 1), thereby matching the levels observed for equities. Values (see Figure 1) above $8$ and $10$ are also common and correspond to annualized volatilities of $54.5$ and $148.4$, which are more common for a highly volatile cryptocurrency, e.g., Bitcoin. Not only the highest realized variance but also the highest level of attention toward the ruble is found on the $28^{th}$ of February 2022, i.e., at $12.19$ in $ln RV$, which corresponds to $443.6$ in annualized form; these changes occurred during the day that the Central Bank of Russia increased the key interest rate to 20\%.

\input{tables/table_1}

\subsection{Explaining price fluctuations}

Our key results for the USD/RUB and EUR/RUB exchange rate intraday price fluctuations are reported in Tables 2 and 3, respectively. We report the results from seven benchmark models that allow us to compare the roles of a variable of interest. We first establish baseline results that show that an intraday seasonality pattern of volatility not only exists (Panel D in Table 1 and 2) but is also non-linear (lowest for the 00:00 to 04:00 and 16:00 to 20:00 UTC trading windows) and considerable in size, i.e., with an elasticity coefficient of up to $-0.20$ for the USD/RUB exchange rate and up to $-0.21$ for the EUR/USD exchange rate. We also control for the crisis week ranging from the $21^{st}$ of February (Monday) to the $28^{th}$ of February (Monday) 2022, during which we observe a tendency for the price variation to systematically increase (Panel E in Tables 1 and 2). The estimated parameters correspond to consecutive days. Thus, the estimated growth over that period of time is chained, which leads to the extreme price variation that has already been observed in Figure 1.

Adding attention measures (Model 2) substantially increases the model's fit ($R^2$ and adjusted $R^2$). Notably, the serial correlation of the residuals is insignificant, which suggests much better specifications than those in Model 1. The ruble-related attention is positive and statistically significant with a considerable effect (elasticity of $0.58$ for the USD/RUB exchange rate and $0.56$ for the EUR/RUB exchange rate) for both exchange rates. The general market attention is relevant, positive, and significant for the USD/RUB exchange rate, which is not that significant given that the search terms are more related to US financial markets. The price fluctuations appear to be higher for the EUR/RUB exchange rate after the observation of greater attention levels toward economic sanctions and the Russian economy.
\input{tables/table_3}

Changes in investors' expectations (i.e., changes in implied volatilities in Model 3 and 4) are not significant for the USD/RUB exchange rate and do not introduce any changes in the previously reported results for Model 1 and Model 2. On the other hand, we observe that the EUR/RUB exchange rate is highly sensitive to changes in investors' expectations, with an elasticity of $1.17$ (Model 3) decreasing only to $0.94$ after the attention variables are included (Model 4). The fit of the model, however, appears to increase only slightly. One explanation for this apparently different outcome of changes in investors' expectations might be that the consequences of the Russo-Ukrainian war will have a much more profound impact on Europe in the short-term, as Europe is geographically closer to the conflict and thus has stronger ties to the Russian energy sector and the Ukrainian economy. 

Finally, in Models 5 to 7, we include the implied volatility directly, with Model 7 also including changes in the implied volatility. Interestingly, these models lead to the highest level of fit, and the attention variables become insignificant for both the USD/RUB and EUR/RUB exchange rates. On the other hand, the implied volatility shows considerable effect, with the elasticity coefficient at $1.05$ ($0.94$ EUR/RUB), $0.94$ ($0.82$ EUR/RUB) and $0.96$ for Model 5, Model 6 and Model 7 ($0.85$ EUR/RUB), respectively. Thus, irrespective of whether we include attention or changes in investors' expectations in the model, it appears that investors' expectations being embedded in intraday implied volatility is followed by realized price variation. Moreover, while implied volatility already encompasses intraday attention, the increased fit suggests that intraday implied volatility also includes other relevant pricing factors that are not captured by attention alone. As before, we observe that changes in investors' expectations are more relevant for the EUR/RUB exchange rate, as the elasticity is found to be $1.65$ and highly significant.

\input{tables/table_4}

\subsection{Alternative modeling choices: robustness checks}
It is likely that our results are driven mostly by observations made from January to March 2022, i.e., during the onset of the crisis. We thus re-estimate all the models using a sample taken from the $19^{th}$ of January 2022, which still included $203$ observations per regression. The conclusion drawn previously remain the same\footnote{All the results from this section are available upon request.} In our study, we use the log of the attention measure; however, alternatives are possible \citep[e.g.][]{audrino2020impact}. As attention measure tends to be noisy in the short-term, we use the log of the moving average found over the past six observations. Given our sampling of $4$-hours, this approach corresponds to one-day moving average. In this scenario, not all the attention variables (in Models 2 and 4) are significant, as only attention toward the Russian ruble remains positive and significant. We also considered estimation of intraday volatility within a $2$, $3$ hours and $6$ hours; however, the main reported conclusions do not change much in this situation either. Finally, we also introduced the average volatility over the past five periods into our models, $ln \bar{V}_{(5),t-1} = ln\left( 5^{-1} \sum_{j=1}^{5} V_{t-j}\right)$. For both exchange rate models, the five-period average volatility was positive and statistically significant in Models 1 to 4, but not after the implied volatility was introduced in Models 5 to 7. The main results, the role of attention and implied volatility, remained largely unchanged.

\section{Conclusion}

During fierce periods of market volatility, it is crucial to react fast and adjust to the current market conditions. Such an unstable environment has occurred on the foreign exchange market during the Russo-Ukrainian war in early 2022. We have estimated several volatility models that can describe intraday behavior of the ruble against major currencies, namely, the U.S. dollar and the euro, from the $1^{st}$ of December 2021 to the end of the $7^{th}$ of March 2022. Our models are motivated by the limited attention theory of \citet{barber2008all} and the implied volatility literature on investors' expectations \citep[e.g.][]{plihal2021modeling}. We use implied volatility obtained from the Bloomberg terminal, which represents the expectations of informed institutional investors. On the other hand, we extract attention measures from Google Trends, which is a broader approach that captures not only the attention of the general population but also that of individual investors. Existing literature uses daily or lower frequency measures of attention and investor's expectations. Part of our contribution lies in the fact that we use intraday attention and implied volatilities and we show that despite the increased sampling frequency, both can be useful.

We provide early evidence on the effects of the imposed economic sanctions and the conflict on the relevant foreign exchange market, thus contributing to this new emerging field \citep[e.g.][]{mamonov2021sorry, ozili2022global, liadze2022economic, halouskova2022}. Specifically, we find that more attention being paid toward: i) the financial market, ii) the Russian ruble and iii) the relationship between economic sanctions and the Russian economy tends to precede larger price fluctuations on the foreign exchange market. However, investors' expectations, as given by implied volatility, seem to encompass information from attention measures and even beyond. These results are similar for both the USD/RUB and EUR/RUB exchange rates, although we find that changing investors' expectations (i.e., changes in implied volatilities) are relevant for the euro as well, which indicates the increased sensitivity of this currency pair to short-term changes in expectations. The geographical position of the Euro zone is likely behind this effect, as it is much closer to the epicenter of the conflict compared to the United States.

While attention measures are helpful for explaining future price changes, from a practical point of view, their timely implementation is challenging, as search terms are not known in advance, specifically in advance of a yet unknown event. This is not the case for implied volatilities. Our results indicate that during unexpected crisis periods, such as the Russo-Ukrainian war, an accurate model of future price fluctuations, should utilize high-frequency quoted implied volatility measures. Such data are available in real-time, their implementation into existing models is straightforward, while they also appear to encompass the information content of attention measures.

\newpage
\appendix
\section{}

\input{tables/table_5}

\newpage
\section*{Acknowledgments}
The authors would like to thank the participants of the MUES research seminars, particularly Martina Halousková, Matůš Horváth, and Daniel Stašek, for their constructive comments. Moreover, the authors highly appreciate feedback from Tomáš Výrost, Zuzana Košťálová, and Oleg Deev.

This research was supported by the Czech Science Foundation (GACR), nr. 22-27075S.

{\scriptsize
\setstretch{1}
\setlength{\bibsep}{1pt}
\bibliography{mybibfile}
}

\end{document}

%% file: tables/table_1.tex
\begin{table}[t]
\centering
\setlength{\tabcolsep}{2.3pt}
\renewcommand{\arraystretch}{1.1} 
\tiny

\caption{Data characteristics: realized and implied measures of price fluctuations and global attention measures}
\label{table_1}

\begin{tabular}{cccccccccccccc}
\toprule
\textbf{Terms}  & \textbf{Mean} & \textbf{SD} & \textbf{Min} & \textbf{Date} & \textbf{Hour} & \textbf{Max} & \textbf{Date} & \textbf{Hour} & \textbf{Skew} & \textbf{Kurt} & \textbf{$\rho(1)$} & \textbf{$\rho(6)$} & \textbf{$\rho(30)$} \\ \midrule
\multicolumn{14}{l}{\textbf{Panel A: Price   fluctuations on USD to RUB}}                                                                                                                                                                            \\
$ln RV_t$              & 5.88          & 1.86        & 2.64         & 22-Dec-21     & 20:00:00      & 12.19        & 28-Feb-22     & 8:00:00       & 1.36          & 4.88          & 0.75               & 0.70               & 0.34                \\
$ln IV_t$              & 6.00          & 1.07        & 4.91         & 17-Dec-21     & 12:00:00      & 9.23         & 4-Mar-22      & 20:00:00      & 1.82          & 5.52          & 0.99               & 0.91               & 0.53                \\
$\Delta ln IV_t$      & 0.01          & 0.08        & -0.23        & 24-Feb-22     & 4:00:00       & 1.27         & 24-Feb-22     & 20:00:00      & 9.57          & 139.42        & 0.31               & -0.09              & 0.07                \\
 \midrule
\multicolumn{14}{l}{\textbf{Panel B: Price   fluctuations on EUR to RUB}}                                                                                                                                                                            \\
$ln RV_t$               & 5.90          & 1.83        & 2.90         & 23-Dec-21     & 20:00:00      & 12.18        & 28-Feb-22     & 8:00:00       & 1.44          & 5.03          & 0.76               & 0.69               & 0.32                \\
$ln IV_t$             & 5.94          & 1.02        & 4.99         & 17-Dec-21     & 12:00:00      & 9.10         & 4-Mar-22      & 12:00:00      & 1.92          & 5.97          & 0.99               & 0.91               & 0.51                \\
$\Delta ln IV_t$      & 0.01          & 0.06        & -0.15        & 25-Feb-22     & 16:00:00      & 0.71         & 28-Feb-22     & 4:00:00       & 6.64          & 73.10         & 0.41               & -0.01              & 0.17                \\
 \midrule
\multicolumn{14}{l}{\textbf{Panel C: Attention   measures}}                                                                                                                                                                                          \\
General market ($ln G_t$)  & 3.16          & 0.61        & 2.05         & 13-Dec-21     & 4:00:00       & 5.00         & 24-Feb-22     & 16:00:00      & 0.49          & 2.61          & 0.79               & 0.85               & 0.54                \\
Ruble ($ln R_t$)           & 1.28          & 1.17        & 0.15         & 4-Jan-22      & 4:00:00       & 5.92         & 28-Feb-22     & 8:00:00       & 1.77          & 5.51          & 0.92               & 0.86               & 0.48                \\
Russian economy ($ln E_t$) & 1.96          & 0.79        & 0.46         & 21-Dec-21     & 4:00:00       & 4.18         & 24-Feb-22     & 20:00:00      & 0.82          & 3.45          & 0.66               & 0.61               & 0.31                \\ 
\bottomrule
\multicolumn{14}{l}{\tiny{\parbox{0.85\linewidth}{\vspace{2pt}\textit{
Notes: SD denotes the standard deviation, Skew and Kurt skewness and kurtosis, while $\rho(.)$ autocorrelation coefficient at the given lag order. Given the high-persistence of the implied volatility series we run a panel-unit root test, where with the monthly implied volatility (IV) is complemented with daily, 1 week, 2 week, and 3 month implied volatility measures and consequently a null hypothesis that all series are stationary is run via the test of \citet{pesaran2007simple}, that accounts for the cross-sectional dependence in multiple implied volatility measures. The same approach is employed for first differences of implied volatilities, the three attention measures (within one panel) and two realized volatility measures (again within one panel). All results reject the null that all series are stationary.
}}}}  
\end{tabular}
\end{table}

%% file: tables/table_3.tex
\begin{table}[t]
\centering
\setlength{\tabcolsep}{2pt}
\renewcommand{\arraystretch}{1} 
\scriptsize

\caption{Volatility model for the USD/RUB exchange pair}
\label{table_3}

\begin{tabular}{lccccccc}
\toprule
                                    & \textbf{Model 1} & \textbf{Model 2} & \textbf{Model 3} & \textbf{Model 4} & \textbf{Model 5} & \textbf{Model 6} & \textbf{Model 7} \\
                                    \midrule
\multicolumn{8}{l}{Panel A: Auto-regressive term}                                                                                                                                                                           \\
Constant                            & \textbf{1.60}      $^c$    & \textbf{1.55}      $^c$    & \textbf{1.60}      $^c$    & \textbf{1.56} $^c$    & \textbf{-1.78} $^c$    & \textbf{-1.67} $^b$    & \textbf{-1.68} $^b$ \\
$ln RV_{t-1}$                          & \textbf{0.81}      $^c$    & \textbf{0.37}      $^c$    & \textbf{0.81}      $^c$    & \textbf{0.37} $^c$    & \textbf{0.27}  $^c$    & \textbf{0.26}  $^c$    & \textbf{0.25}  $^c$ \\
\midrule
\multicolumn{8}{l}{Panel B: Implied volatility measures}                                                                                                                                                                \\
$\Delta ln IV_{t-1}$                   &                            &                            & 0.27                       & 0.08                  &                        &                        & \textbf{0.61}  $^a$ \\
$ln IV_{t-1}$                          &                            &                            &                            &                       & \textbf{1.05}  $^c$    & \textbf{0.94}  $^c$    & \textbf{0.96}  $^c$ \\
\midrule
\multicolumn{8}{l}{Panel C: Attention   measures}                                                                                                                                                                         \\
$ln G_{t-1}$ (General market)         &                            & \textbf{0.40}   $^b$       &                            & \textbf{0.40} $^b$    &                        & 0.15                   & 0.13                \\
$ln R_{t-1}$ (Ruble related)          &                            & \textbf{0.58}   $^b$       &                            & \textbf{0.58} $^b$    &                        & 0.04                   & 0.04                \\
$ln E_{t-1}$ (Sanctions \&   Economy) &                            & \textbf{0.16}   $^a$       &                            & \textbf{0.16} $^a$    &                        & 0.03                   & 0.03                \\
\midrule
\multicolumn{8}{l}{Panel D: Control variables - intraday seasonality}                                                                                                                                                   \\
$ln RV_{t-1} \times I_t(j=1)$          & \textbf{-0.16}     $^c$    & \textbf{-0.10}     $^c$    & \textbf{-0.16}     $^c$    & \textbf{-0.10}$^c$    & \textbf{-0.09} $^c$    & \textbf{-0.09} $^c$    & \textbf{-0.09} $^c$ \\
$ln RV_{t-1} \times I_t(j=2)$          & -0.05                      & 0.02                       & -0.05                      & 0.02                  & -0.01                  & 0.00                   & 0.00                \\
$ln RV_{t-1} \times I_t(j=3)$          & \textbf{-0.08}     $^b$    & -0.01                      & \textbf{-0.08}     $^b$    & -0.01                 & -0.03                  & -0.02                  & -0.02               \\
$ln RV_{t-1} \times I_t(j=4)$          & -0.03                      & 0.03                       & -0.04                      & 0.03                  & 0.01                   & 0.03                   & 0.02                \\
$ln RV_{t-1} \times I_t(j=5)$          & \textbf{-0.20}     $^c$    & \textbf{-0.15}     $^c$    & \textbf{-0.20}     $^c$    & \textbf{-0.15}$^c$    & \textbf{-0.13} $^c$    & \textbf{-0.13} $^c$    & \textbf{-0.13} $^c$ \\
\midrule
\multicolumn{8}{l}{Panel E: Control variables - onset of the war}                                                                                                                                                       \\
$ln RV_{t-1} \times I_t(q = 21 Feb)$     & 0.03                       & \textbf{0.03}      $^a$    & 0.03                       & \textbf{0.03} $^a$    & 0.02                   & \textbf{0.02}  $^a$    & 0.02                \\
$ln RV_{t-1} \times I_t(q = 22 Feb)$   & 0.04                       & -0.04                      & 0.04                       & -0.04                 & -0.02                  & -0.02                  & -0.02               \\
$ln RV_{t-1} \times I_t(q = 23 Feb)$   & \textbf{0.07}      $^c$    & -0.02                      & \textbf{0.07}      $^c$    & -0.02                 & \textbf{-0.06} $^b$    & \textbf{-0.06} $^b$    & \textbf{-0.06} $^b$ \\
$ln RV_{t-1} \times I_t(q = 24 Feb)$   & \textbf{0.10}      $^b$    & -0.02                      & 0.10                       & -0.02                 & \textbf{0.09}  $^b$    & \textbf{0.07}  $^a$    & \textbf{0.07}  $^a$ \\
$ln RV_{t-1} \times I_t(q = 25 Feb)$   & \textbf{0.13}      $^c$    & -0.02                      & \textbf{0.13}      $^c$    & -0.02                 & \textbf{0.05}  $^a$    & 0.03                   & 0.04                \\
$ln RV_{t-1} \times I_t(q = 28 Feb)$   & \textbf{0.13}      $^b$    & \textbf{0.05}      $^a$    & \textbf{0.12}      $^b$    & \textbf{0.05} $^a$    & \textbf{0.14}  $^b$    & \textbf{0.13}  $^b$    & \textbf{0.11}  $^b$ \\
\midrule
\multicolumn{8}{l}{Panel F: Model diagnostics}                                                                                                                                                                          \\
Serial dependence (pval)            & 0.01                       & 0.42                       & 0.01                       & 0.42                  & 0.50                   & 0.53                   & 0.55                \\
$R^2$                               & 64.9\%                     & 72.0\%                     & 64.9\%                     & 72.0\%                & 74.1\%                 & 74.1\%                 & 74.2\%              \\
adjusted $R^2$                      & 63.7\%                     & 70.9\%                     & 63.7\%                     & 70.8\%                & 73.2\%                 & 73.0\%                 & 73.0\%              \\ 
\bottomrule
\multicolumn{8}{l}{\tiny{\parbox{0.65\linewidth}{\vspace{2pt}\textit{
Notes: a. b. c denote 10\%. 5\%. and 1\% significance level of the estimated parameter. Serial dependence of residuals is established via the \citet{escanciano2009automatic} test with maximum lag set to 30 (one week).
}}}}  
\end{tabular}
\end{table}

%% file: tables/table_4.tex
\begin{table}[t]
\centering
\setlength{\tabcolsep}{2pt}
\renewcommand{\arraystretch}{1} 
\scriptsize

\caption{Volatility model for the EUR/RUB exchange pair}
\label{table_4}

\begin{tabular}{lccccccc}
\toprule
                                    & \textbf{Model 1} & \textbf{Model 2} & \textbf{Model 3} & \textbf{Model 4} & \textbf{Model 5} & \textbf{Model 6} & \textbf{Model 7} \\
                                    \midrule
\multicolumn{8}{l}{Panel A: Auto-regressive term}                                                                                                                                                       \\
Constant                            & \textbf{1.40}  $^c$ & \textbf{1.85}  $^c$   & \textbf{1.42}   $^c$ & \textbf{1.90}  $^c$   & \textbf{-1.57}  $^b$ & \textbf{-1.06}      & \textbf{-1.07} $^a$ \\
$ln RV_{t-1}$                         & \textbf{0.85}  $^c$ & \textbf{0.47}  $^c$   & \textbf{0.85}   $^c$ & \textbf{0.47}  $^c$   & \textbf{0.38}   $^c$ & \textbf{0.38}  $^c$ & \textbf{0.37}  $^c$ \\
\midrule
\multicolumn{8}{l}{Panel B: Implied   volatility measures}                                                                                                                                              \\
$\Delta ln IV_{t-1}$                    &                    &                        & \textbf{1.17}  $^c$ & \textbf{0.94} $^a$   &                      &                         & \textbf{1.65} $^c$ \\
$ln IV_{t-1}$                          &                    &                        &                      &                       & \textbf{0.94}   $^c$ &   \textbf{0.82} $^c$   & \textbf{0.85} $^c$ \\
\midrule
\multicolumn{8}{l}{Panel C: Attention   measures}                                                                                                                                                       \\
$ln G_{t-1}$ (General market)         &                   & \textbf{0.18} $^a$ &      & 0.16                &      & -0.01                                                    & -0.05               \\
$ln R_{t-1}$ (Ruble related)          &                   & \textbf{1.56} $^b$ &      & \textbf{0.56}  $^b$ &      & 0.09                                                    & 0.08               \\
$ln E_{t-1}$ (Sanctions \&   Economy) &                   & \textbf{0.18} $^b$ &      & \textbf{0.18}  $^b$ &      & 0.07                                                    & 0.07               \\
\midrule
\multicolumn{8}{l}{Panel D: Control   variables - intraday seasonality}                                                                                                                                 \\
$ln RV_{t-1} \times I_t(j=1)$          & \textbf{-0.21} $^c$ & \textbf{-0.16} $^c$   & \textbf{-0.21}  $^c$ & \textbf{-0.16} $^c$   & \textbf{-0.14}  $^c$ & \textbf{-0.14} $^c$ & \textbf{-0.14} $^c$ \\
$ln RV_{t-1} \times I_t(j=2)$          & -0.02               & 0.018                 & -0.02                & 0.018                 & -0.01                & -0.01               & -0                  \\
$ln RV_{t-1} \times I_t(j=3)$          & \textbf{-0.08} $^b$ & -0.04                 & \textbf{-0.09}  $^b$ & -0.04                 & -0.04                & -0.04               & -0.05               \\
$ln RV_{t-1} \times I_t(j=4)$          & -0.05               & -0.01                 & -0.05                & -0.01                 & -0.01                & -0.01               & -0.01               \\
$ln RV_{t-1} \times I_t(j=5)$          & \textbf{-0.19} $^c$ & \textbf{-0.15} $^c$   & \textbf{-0.19}  $^c$ & \textbf{-0.15} $^c$   & \textbf{-0.14}  $^c$ & \textbf{-0.14} $^c$ & \textbf{-0.14} $^c$ \\
\midrule
\multicolumn{8}{l}{Panel E: Control   variables - onset of the war}                                                                                                                                     \\
$ln RV_{t-1} \times I_t(q = 21 Feb)$     & 0.03                & \textbf{0.03}  $^a$   & 0.01                 & 0.02                 & \textbf{0.03}   $^a$ & \textbf{0.03}  $^a$ & 0.02                \\
$ln RV_{t-1} \times I_t(q = 22 Feb)$   & 0.00                & -0.05                 & 0.00                 & -0.06                & -0.02                & -0.02               & -0.03               \\
$ln RV_{t-1} \times I_t(q = 23 Feb)$   & \textbf{0.05}  $^b$ & -0.05                 & \textbf{0.05}   $^b$ & -0.05                & \textbf{-0.10}  $^c$ & \textbf{-0.09} $^c$ & \textbf{-0.09} $^c$ \\
$ln RV_{t-1} \times I_t(q = 24 Feb)$   & \textbf{0.07}  $^b$ & -0.02                 & \textbf{0.06}   $^a$ & -0.03                & \textbf{0.08}   $^b$ & \textbf{0.06}  $^a$ & 0.04                \\
$ln RV_{t-1} \times I_t(q = 25 Feb)$   & \textbf{0.13}  $^c$ & -0.01                 & \textbf{0.14}   $^c$ & 0.00                 & \textbf{0.05}   $^b$ & 0.03                & \textbf{0.05}  $^a$ \\
$ln RV_{t-1} \times I_t(q = 28 Feb)$   & \textbf{0.09}  $^b$ & 0.02                  & \textbf{0.06}   $^b$ & 0.00                 & \textbf{0.12}   $^b$ & \textbf{0.11}  $^b$ & \textbf{0.07}  $^a$ \\
\midrule
\multicolumn{8}{l}{Panel F: Model   diagnostics}                                                                                                                                                        \\
Serial dependence (pval)            & 0.00               & 0.29                  & 0.00              & 0.27              & 0.54              & 0.57              & 0.57             \\
$R^2$                               & 68.9\%             & 74.2\%                & 69.0\%            & 74.3\%            & 75.8\%            & 75.8\%            & 76.0\%           \\
adjusted $R^2$                      & 67.9\%             & 73.2\%                & 67.9\%            & 73.2\%            & 75.0\%            & 74.8\%            & 74.9\%           \\ 
\bottomrule
\multicolumn{8}{l}{\tiny{\parbox{0.65\linewidth}{\vspace{2pt}\textit{
Notes: a. b. c denote 10\%. 5\%. and 1\% significance level of the estimated parameter. Serial dependence of residuals is established via the \citet{escanciano2009automatic} test with maximum lag set to 30 (one week).
}}}}  
\end{tabular}
\end{table}

%% file: tables/table_5.tex
\begin{table}[h]
\centering
\setlength{\tabcolsep}{5pt}
\renewcommand{\arraystretch}{1.2} 
\footnotesize

\caption{Search Terms}
\label{table_5}

\begin{tabular}{lll}
\toprule
Ruble                 & General financial market & Russian economy          \\
\midrule
ruble                 & S\&P 500                 & economic sanctions       \\
usd rub               & VIX                      & Vnesheconombank          \\
eur rub               & stock market             & Promsvyazbank            \\
btc rub               & FX market                & VTB Bank                 \\
Russian central bank  & dow jones                & Sberbank                 \\
Russian interest rate & nasdaq                   & Alfa Bank                \\
                      & nyse                     & Rossiya Bank             \\
                      &                          & SWFIT Russia             \\
                      &                          & asset freeze             \\
                      &                          & Nord Stream 2            \\
                      &                          & export controls          \\
                      &                          & Moscow stock exchange    \\
                      &                          & currency control         \\
                      &                          & fx reserves              \\
                      &                          & Maersk Russia            \\
                      &                          & British Petroleum Russia \\
                      &                          & Ikea Russia              \\
                      &                          & Apple Russia             \\
                      &                          & Disney Russia            \\
                      &                          & Equinor Russia           \\
                      &                          & Exxon Russia             \\
                      &                          & Shell Russia             \\
                      &                          & Mastercard Russia        \\
                      &                          & Boeing Russia            \\
                      &                          & Airbus Russia            \\
                      &                          & American Express Russia  \\
                      &                          & Dell Russia              \\
                      &                          & Ford Russia              \\
                      &                          & Google Russia            \\
                      &                          & Airbnb Russia            \\
                      &                          & Meta Russia              \\
                      &                          & HM Russia                \\
                      &                          & McDonalds Russia         \\
                      &                          & Nike Russia              \\
                      &                          & Visa Russia             \\
                      \bottomrule
\end{tabular}
\end{table}

%% file: main.bbl
\begin{thebibliography}{46}
\expandafter\ifx\csname natexlab\endcsname\relax\def\natexlab#1{#1}\fi
\providecommand{\url}[1]{\texttt{#1}}
\providecommand{\href}[2]{#2}
\providecommand{\path}[1]{#1}
\providecommand{\DOIprefix}{doi:}
\providecommand{\ArXivprefix}{arXiv:}
\providecommand{\URLprefix}{URL: }
\providecommand{\Pubmedprefix}{pmid:}
\providecommand{\doi}[1]{\href{http://dx.doi.org/#1}{\path{#1}}}
\providecommand{\Pubmed}[1]{\href{pmid:#1}{\path{#1}}}
\providecommand{\bibinfo}[2]{#2}
\ifx\xfnm\relax \def\xfnm[#1]{\unskip,\space#1}\fi
\bibitem[{Andersen et~al.(2001)Andersen, Bollerslev, Diebold and
  Labys}]{andersen2001distribution}
\bibinfo{author}{Andersen, T.G.}, \bibinfo{author}{Bollerslev, T.},
  \bibinfo{author}{Diebold, F.X.}, \bibinfo{author}{Labys, P.},
  \bibinfo{year}{2001}.
\newblock \bibinfo{title}{The distribution of realized exchange rate
  volatility}.
\newblock \bibinfo{journal}{Journal of the American statistical association}
  \bibinfo{volume}{96}, \bibinfo{pages}{42--55}.
\bibitem[{Aslanidis et~al.(2022)Aslanidis, Bariviera and
  L{\'o}pez}]{aslanidis2022link}
\bibinfo{author}{Aslanidis, N.}, \bibinfo{author}{Bariviera, A.F.},
  \bibinfo{author}{L{\'o}pez, {\'O}.G.}, \bibinfo{year}{2022}.
\newblock \bibinfo{title}{The link between cryptocurrencies and google trends
  attention}.
\newblock \bibinfo{journal}{Finance Research Letters} ,
  \bibinfo{pages}{102654}.
\bibitem[{Audrino et~al.(2020)Audrino, Sigrist and
  Ballinari}]{audrino2020impact}
\bibinfo{author}{Audrino, F.}, \bibinfo{author}{Sigrist, F.},
  \bibinfo{author}{Ballinari, D.}, \bibinfo{year}{2020}.
\newblock \bibinfo{title}{The impact of sentiment and attention measures on
  stock market volatility}.
\newblock \bibinfo{journal}{International Journal of Forecasting}
  \bibinfo{volume}{36}, \bibinfo{pages}{334--357}.
\bibitem[{Barber and Odean(2008)}]{barber2008all}
\bibinfo{author}{Barber, B.M.}, \bibinfo{author}{Odean, T.},
  \bibinfo{year}{2008}.
\newblock \bibinfo{title}{All that glitters: The effect of attention and news
  on the buying behavior of individual and institutional investors}.
\newblock \bibinfo{journal}{The review of financial studies}
  \bibinfo{volume}{21}, \bibinfo{pages}{785--818}.
\bibitem[{Bleher and Dimpfl(2019)}]{bleher2019today}
\bibinfo{author}{Bleher, J.}, \bibinfo{author}{Dimpfl, T.},
  \bibinfo{year}{2019}.
\newblock \bibinfo{title}{Today i got a million, tomorrow, i don't know: on the
  predictability of cryptocurrencies by means of google search volume}.
\newblock \bibinfo{journal}{International Review of Financial Analysis}
  \bibinfo{volume}{63}, \bibinfo{pages}{147--159}.
\bibitem[{Bleher and Dimpfl(2021)}]{bleher2021knitting}
\bibinfo{author}{Bleher, J.}, \bibinfo{author}{Dimpfl, T.},
  \bibinfo{year}{2021}.
\newblock \bibinfo{title}{Knitting multi-annual high-frequency google trends to
  predict inflation and consumption.}
\newblock \bibinfo{journal}{Econometrics and Statistics} .
\bibitem[{Bollerslev(1986)}]{bollerslev1986generalized}
\bibinfo{author}{Bollerslev, T.}, \bibinfo{year}{1986}.
\newblock \bibinfo{title}{Generalized autoregressive conditional
  heteroskedasticity}.
\newblock \bibinfo{journal}{Journal of econometrics} \bibinfo{volume}{31},
  \bibinfo{pages}{307--327}.
\bibitem[{Bollerslev et~al.(2018)Bollerslev, Hood, Huss and
  Pedersen}]{bollerslev2018risk}
\bibinfo{author}{Bollerslev, T.}, \bibinfo{author}{Hood, B.},
  \bibinfo{author}{Huss, J.}, \bibinfo{author}{Pedersen, L.H.},
  \bibinfo{year}{2018}.
\newblock \bibinfo{title}{Risk everywhere: Modeling and managing volatility}.
\newblock \bibinfo{journal}{The Review of Financial Studies}
  \bibinfo{volume}{31}, \bibinfo{pages}{2729--2773}.
\bibitem[{Canina and Figlewski(1993)}]{canina1993informational}
\bibinfo{author}{Canina, L.}, \bibinfo{author}{Figlewski, S.},
  \bibinfo{year}{1993}.
\newblock \bibinfo{title}{The informational content of implied volatility}.
\newblock \bibinfo{journal}{The Review of Financial Studies}
  \bibinfo{volume}{6}, \bibinfo{pages}{659--681}.
\bibitem[{Chan(2020)}]{chan2020temporal}
\bibinfo{author}{Chan, W.S.}, \bibinfo{year}{2020}.
\newblock \bibinfo{title}{On temporal aggregation of some nonlinear time-series
  models}.
\newblock \bibinfo{journal}{Econometrics and Statistics} .
\bibitem[{Chang and Taylor(2003)}]{chang2003information}
\bibinfo{author}{Chang, Y.}, \bibinfo{author}{Taylor, S.J.},
  \bibinfo{year}{2003}.
\newblock \bibinfo{title}{Information arrivals and intraday exchange rate
  volatility}.
\newblock \bibinfo{journal}{Journal of International Financial Markets,
  Institutions and Money} \bibinfo{volume}{13}, \bibinfo{pages}{85--112}.
\bibitem[{Chronopoulos et~al.(2018)Chronopoulos, Papadimitriou and
  Vlastakis}]{chronopoulos2018information}
\bibinfo{author}{Chronopoulos, D.K.}, \bibinfo{author}{Papadimitriou, F.I.},
  \bibinfo{author}{Vlastakis, N.}, \bibinfo{year}{2018}.
\newblock \bibinfo{title}{Information demand and stock return predictability}.
\newblock \bibinfo{journal}{Journal of International Money and Finance}
  \bibinfo{volume}{80}, \bibinfo{pages}{59--74}.
\bibitem[{Corsi(2009)}]{corsi2009simple}
\bibinfo{author}{Corsi, F.}, \bibinfo{year}{2009}.
\newblock \bibinfo{title}{A simple approximate long-memory model of realized
  volatility}.
\newblock \bibinfo{journal}{Journal of Financial Econometrics}
  \bibinfo{volume}{7}, \bibinfo{pages}{174--196}.
\bibitem[{Da et~al.(2011)Da, Engelberg and Gao}]{da2011search}
\bibinfo{author}{Da, Z.}, \bibinfo{author}{Engelberg, J.},
  \bibinfo{author}{Gao, P.}, \bibinfo{year}{2011}.
\newblock \bibinfo{title}{In search of attention}.
\newblock \bibinfo{journal}{The journal of finance} \bibinfo{volume}{66},
  \bibinfo{pages}{1461--1499}.
\bibitem[{Dimpfl and Jank(2016)}]{dimpfl2016can}
\bibinfo{author}{Dimpfl, T.}, \bibinfo{author}{Jank, S.}, \bibinfo{year}{2016}.
\newblock \bibinfo{title}{Can internet search queries help to predict stock
  market volatility?}
\newblock \bibinfo{journal}{European Financial Management}
  \bibinfo{volume}{22}, \bibinfo{pages}{171--192}.
\bibitem[{Ederington and Lee(2001)}]{ederington2001intraday}
\bibinfo{author}{Ederington, L.}, \bibinfo{author}{Lee, J.H.},
  \bibinfo{year}{2001}.
\newblock \bibinfo{title}{Intraday volatility in interest-rate and
  foreign-exchange markets: Arch, announcement, and seasonality effects}.
\newblock \bibinfo{journal}{Journal of Futures Markets: Futures, Options, and
  Other Derivative Products} \bibinfo{volume}{21}, \bibinfo{pages}{517--552}.
\bibitem[{Escanciano and Lobato(2009)}]{escanciano2009automatic}
\bibinfo{author}{Escanciano, J.C.}, \bibinfo{author}{Lobato, I.N.},
  \bibinfo{year}{2009}.
\newblock \bibinfo{title}{An automatic portmanteau test for serial
  correlation}.
\newblock \bibinfo{journal}{Journal of Econometrics} \bibinfo{volume}{151},
  \bibinfo{pages}{140--149}.
\bibitem[{Fassas and Siriopoulos(2021)}]{fassas2021implied}
\bibinfo{author}{Fassas, A.P.}, \bibinfo{author}{Siriopoulos, C.},
  \bibinfo{year}{2021}.
\newblock \bibinfo{title}{Implied volatility indices--a review}.
\newblock \bibinfo{journal}{The Quarterly Review of Economics and Finance}
  \bibinfo{volume}{79}, \bibinfo{pages}{303--329}.
\bibitem[{Gkillas et~al.(2021)Gkillas, Konstantatos and
  Siriopoulos}]{gkillas2021uncertainty}
\bibinfo{author}{Gkillas, K.}, \bibinfo{author}{Konstantatos, C.},
  \bibinfo{author}{Siriopoulos, C.}, \bibinfo{year}{2021}.
\newblock \bibinfo{title}{Uncertainty due to infectious diseases and
  stock--bond correlation}.
\newblock \bibinfo{journal}{Econometrics} \bibinfo{volume}{9},
  \bibinfo{pages}{17}.
\bibitem[{Goddard et~al.(2015)Goddard, Kita and Wang}]{goddard2015investor}
\bibinfo{author}{Goddard, J.}, \bibinfo{author}{Kita, A.},
  \bibinfo{author}{Wang, Q.}, \bibinfo{year}{2015}.
\newblock \bibinfo{title}{Investor attention and fx market volatility}.
\newblock \bibinfo{journal}{Journal of International Financial Markets,
  Institutions and Money} \bibinfo{volume}{38}, \bibinfo{pages}{79--96}.
\bibitem[{Gonzalez-Perez(2015)}]{gonzalez2015model}
\bibinfo{author}{Gonzalez-Perez, M.T.}, \bibinfo{year}{2015}.
\newblock \bibinfo{title}{Model-free volatility indexes in the financial
  literature: A review}.
\newblock \bibinfo{journal}{International Review of Economics \& Finance}
  \bibinfo{volume}{40}, \bibinfo{pages}{141--159}.
\bibitem[{Halousková et~al.(2022)Halousková, Horváth and
  Daniel}]{halouskova2022}
\bibinfo{author}{Halousková, M.}, \bibinfo{author}{Horváth, M.},
  \bibinfo{author}{Daniel, S.}, \bibinfo{year}{2022}.
\newblock \bibinfo{title}{The role of investor attention in global assets price
  variation during the war in ukraine (working paper)}.
\newblock \bibinfo{journal}{Available at http://arxiv.org/abs/2205.05985} .
\bibitem[{Kim et~al.(2019)Kim, Lu{\v{c}}ivjansk{\'a}, Moln{\'a}r and
  Villa}]{kim2019google}
\bibinfo{author}{Kim, N.}, \bibinfo{author}{Lu{\v{c}}ivjansk{\'a}, K.},
  \bibinfo{author}{Moln{\'a}r, P.}, \bibinfo{author}{Villa, R.},
  \bibinfo{year}{2019}.
\newblock \bibinfo{title}{Google searches and stock market activity: Evidence
  from norway}.
\newblock \bibinfo{journal}{Finance Research Letters} \bibinfo{volume}{28},
  \bibinfo{pages}{208--220}.
\bibitem[{Kristoufek(2015)}]{kristoufek2015main}
\bibinfo{author}{Kristoufek, L.}, \bibinfo{year}{2015}.
\newblock \bibinfo{title}{What are the main drivers of the bitcoin price?
  evidence from wavelet coherence analysis}.
\newblock \bibinfo{journal}{PloS one} \bibinfo{volume}{10},
  \bibinfo{pages}{e0123923}.
\bibitem[{Liadze et~al.(2022)Liadze, Macchiarelli, Mortimer-Lee and
  Juanino}]{liadze2022economic}
\bibinfo{author}{Liadze, I.}, \bibinfo{author}{Macchiarelli, C.},
  \bibinfo{author}{Mortimer-Lee, P.}, \bibinfo{author}{Juanino, P.S.},
  \bibinfo{year}{2022}.
\newblock \bibinfo{title}{The economic costs of the russia-ukraine conflict}.
\newblock \bibinfo{journal}{Available at SSRN} .
\bibitem[{Liu et~al.(2015)Liu, Patton and Sheppard}]{liu2015does}
\bibinfo{author}{Liu, L.Y.}, \bibinfo{author}{Patton, A.J.},
  \bibinfo{author}{Sheppard, K.}, \bibinfo{year}{2015}.
\newblock \bibinfo{title}{Does anything beat 5-minute rv? a comparison of
  realized measures across multiple asset classes}.
\newblock \bibinfo{journal}{Journal of Econometrics} \bibinfo{volume}{187},
  \bibinfo{pages}{293--311}.
\bibitem[{Ly{\'o}csa et~al.(2020a)Ly{\'o}csa, Baum{\"o}hl, V{\`y}rost and
  Moln{\'a}r}]{lyocsa2020fear}
\bibinfo{author}{Ly{\'o}csa, {\v{S}}.}, \bibinfo{author}{Baum{\"o}hl, E.},
  \bibinfo{author}{V{\`y}rost, T.}, \bibinfo{author}{Moln{\'a}r, P.},
  \bibinfo{year}{2020}a.
\newblock \bibinfo{title}{Fear of the coronavirus and the stock markets}.
\newblock \bibinfo{journal}{Finance research letters} \bibinfo{volume}{36},
  \bibinfo{pages}{101735}.
\bibitem[{Ly{\'o}csa et~al.(2020b)Ly{\'o}csa, Moln{\'a}r, Pl{\'\i}hal and
  {\v{S}}ira{\v{n}}ov{\'a}}]{lyocsa2020impact}
\bibinfo{author}{Ly{\'o}csa, {\v{S}}.}, \bibinfo{author}{Moln{\'a}r, P.},
  \bibinfo{author}{Pl{\'\i}hal, T.}, \bibinfo{author}{{\v{S}}ira{\v{n}}ov{\'a},
  M.}, \bibinfo{year}{2020}b.
\newblock \bibinfo{title}{Impact of macroeconomic news, regulation and hacking
  exchange markets on the volatility of bitcoin}.
\newblock \bibinfo{journal}{Journal of Economic Dynamics and Control}
  \bibinfo{volume}{119}, \bibinfo{pages}{103980}.
\bibitem[{Mamonov and Pestova(2021)}]{mamonov2021sorry}
\bibinfo{author}{Mamonov, M.}, \bibinfo{author}{Pestova, A.},
  \bibinfo{year}{2021}.
\newblock \bibinfo{title}{Sorry, you're blocked.'economic effects of financial
  sanctions on the russian economy}.
\newblock \bibinfo{journal}{CERGE-EI Working Paper Series} .
\bibitem[{Mankoff(2014)}]{mankoff2014russia}
\bibinfo{author}{Mankoff, J.}, \bibinfo{year}{2014}.
\newblock \bibinfo{title}{Russia's latest land grab: How putin won crimea and
  lost ukraine}.
\newblock \bibinfo{journal}{Foreign Aff.} \bibinfo{volume}{93},
  \bibinfo{pages}{60}.
\bibitem[{Massicotte and Eddelbuettel(2021)}]{massicotte2021gtrendsr}
\bibinfo{author}{Massicotte, P.}, \bibinfo{author}{Eddelbuettel, D.},
  \bibinfo{year}{2021}.
\newblock \bibinfo{title}{gtrends{R}: Perform and display {G}oogle trends
  queries}.
\newblock \bibinfo{journal}{R package version 1.5}
  \bibinfo{volume}{https://CRAN.R-project.org/package=gtrendsR}.
\bibitem[{Melvin and Yin(2000)}]{melvin2000public}
\bibinfo{author}{Melvin, M.}, \bibinfo{author}{Yin, X.}, \bibinfo{year}{2000}.
\newblock \bibinfo{title}{Public information arrival, exchange rate volatility,
  and quote frequency}.
\newblock \bibinfo{journal}{The Economic Journal} \bibinfo{volume}{110},
  \bibinfo{pages}{644--661}.
\bibitem[{Nicholas~Taleb(2021)}]{nicholas2021bitcoin}
\bibinfo{author}{Nicholas~Taleb, N.}, \bibinfo{year}{2021}.
\newblock \bibinfo{title}{Bitcoin, currencies, and fragility}.
\newblock \bibinfo{journal}{Quantitative Finance} \bibinfo{volume}{21},
  \bibinfo{pages}{1249--1255}.
\bibitem[{Ozili(2022)}]{ozili2022global}
\bibinfo{author}{Ozili, P.K.}, \bibinfo{year}{2022}.
\newblock \bibinfo{title}{Global economic consequence of russian invasion of
  ukraine}.
\newblock \bibinfo{journal}{Available at SSRN} .
\bibitem[{Patton et~al.(2009)Patton, Politis and White}]{patton2009correction}
\bibinfo{author}{Patton, A.}, \bibinfo{author}{Politis, D.N.},
  \bibinfo{author}{White, H.}, \bibinfo{year}{2009}.
\newblock \bibinfo{title}{Correction to “automatic block-length selection for
  the dependent bootstrap” by d. politis and h. white}.
\newblock \bibinfo{journal}{Econometric Reviews} \bibinfo{volume}{28},
  \bibinfo{pages}{372--375}.
\bibitem[{Pesaran(2007)}]{pesaran2007simple}
\bibinfo{author}{Pesaran, M.H.}, \bibinfo{year}{2007}.
\newblock \bibinfo{title}{A simple panel unit root test in the presence of
  cross-section dependence}.
\newblock \bibinfo{journal}{Journal of applied econometrics}
  \bibinfo{volume}{22}, \bibinfo{pages}{265--312}.
\bibitem[{Pl{\'\i}hal(2021)}]{plihal2021scheduled}
\bibinfo{author}{Pl{\'\i}hal, T.}, \bibinfo{year}{2021}.
\newblock \bibinfo{title}{Scheduled macroeconomic news announcements and forex
  volatility forecasting}.
\newblock \bibinfo{journal}{Journal of Forecasting} \bibinfo{volume}{40},
  \bibinfo{pages}{1379--1397}.
\bibitem[{Pl{\'\i}hal and Ly{\'o}csa(2021)}]{plihal2021modeling}
\bibinfo{author}{Pl{\'\i}hal, T.}, \bibinfo{author}{Ly{\'o}csa, {\v{S}}.},
  \bibinfo{year}{2021}.
\newblock \bibinfo{title}{Modeling realized volatility of the eur/usd exchange
  rate: Does implied volatility really matter?}
\newblock \bibinfo{journal}{International Review of Economics \& Finance}
  \bibinfo{volume}{71}, \bibinfo{pages}{811--829}.
\bibitem[{Politis and White(2004)}]{politis2004automatic}
\bibinfo{author}{Politis, D.N.}, \bibinfo{author}{White, H.},
  \bibinfo{year}{2004}.
\newblock \bibinfo{title}{Automatic block-length selection for the dependent
  bootstrap}.
\newblock \bibinfo{journal}{Econometric reviews} \bibinfo{volume}{23},
  \bibinfo{pages}{53--70}.
\bibitem[{Polyzos(2022)}]{polyzos2022escalating}
\bibinfo{author}{Polyzos, E.S.}, \bibinfo{year}{2022}.
\newblock \bibinfo{title}{Escalating tension and the war in ukraine: Evidence
  using impulse response functions on economic indicators and twitter
  sentiment}.
\newblock \bibinfo{journal}{Available at SSRN 4058364} .
\bibitem[{Poon and Granger(2003)}]{poon2003forecasting}
\bibinfo{author}{Poon, S.H.}, \bibinfo{author}{Granger, C.W.},
  \bibinfo{year}{2003}.
\newblock \bibinfo{title}{Forecasting volatility in financial markets: A
  review}.
\newblock \bibinfo{journal}{Journal of economic literature}
  \bibinfo{volume}{41}, \bibinfo{pages}{478--539}.
\bibitem[{Racine and Hayfield(2021)}]{Racine2021}
\bibinfo{author}{Racine, J.S.}, \bibinfo{author}{Hayfield, T.},
  \bibinfo{year}{2021}.
\newblock \bibinfo{title}{Nonparametric kernel smoothing methods for mixed data
  types}.
\newblock \bibinfo{journal}{R package version 0.60-11}
  \bibinfo{volume}{https://CRAN.R-project.org/package=np}.
\bibitem[{Rywkin(2014)}]{rywkin2014ukraine}
\bibinfo{author}{Rywkin, M.}, \bibinfo{year}{2014}.
\newblock \bibinfo{title}{Ukraine: between russia and the west}.
\newblock \bibinfo{journal}{American Foreign policy interests}
  \bibinfo{volume}{36}, \bibinfo{pages}{119--126}.
\bibitem[{Samokhvalov(2015)}]{samokhvalov2015ukraine}
\bibinfo{author}{Samokhvalov, V.}, \bibinfo{year}{2015}.
\newblock \bibinfo{title}{Ukraine between russia and the european union:
  triangle revisited}.
\newblock \bibinfo{journal}{Europe-Asia Studies} \bibinfo{volume}{67},
  \bibinfo{pages}{1371--1393}.
\bibitem[{Seemann et~al.(2011)Seemann, McCauley and
  Gunaratne}]{seemann2011intraday}
\bibinfo{author}{Seemann, L.}, \bibinfo{author}{McCauley, J.L.},
  \bibinfo{author}{Gunaratne, G.H.}, \bibinfo{year}{2011}.
\newblock \bibinfo{title}{Intraday volatility and scaling in high frequency
  foreign exchange markets}.
\newblock \bibinfo{journal}{International Review of Financial Analysis}
  \bibinfo{volume}{20}, \bibinfo{pages}{121--126}.
\bibitem[{Shelest(2015)}]{shelest2015after}
\bibinfo{author}{Shelest, H.}, \bibinfo{year}{2015}.
\newblock \bibinfo{title}{After the ukrainian crisis: Is there a place for
  russia?}
\newblock \bibinfo{journal}{Southeast European and Black Sea Studies}
  \bibinfo{volume}{15}, \bibinfo{pages}{191--201}.

\end{thebibliography}
